\documentclass[aps,twocolumn]{revtex4}
\usepackage[colorlinks]{hyperref}
\usepackage[utf8]{inputenc}
\usepackage{mathtools}

\begin{document}

\title{Effects of gravity in extra dimensions  in atomic phenomena}
\author{  V. A. Dzuba, % \orcid{0000-0003-2758-5574},
V. V. Flambaum, % \orcid{0000-0001-8643-7374},  
P. Munro-Laylim% \orcid{0000-0003-0041-3828},
}
\affiliation{School of Physics, University of New South Wales, Sydney 2052, Australia}

\begin{abstract}
We use the difference between theory and experiment for energy intervals in simple atomic systems (hydrogen, muonium, positronium and deuteron) to find limits on the size of extra space dimensions in the Arkani-Hamed  - Dimopoulos - Dvali model for gravitation  potential on short distances. As an additional experimental fact we use absence of the small size gravitational bound states of elementary particles. 
We demonstrate that  the perturbation theory approach does not work and more reliable results are obtained by solving the Dirac equations for an electron in Coulomb and gravitational fields. These results probe smaller distances than distance between nuclei in molecules  and the limits are significantly  stronger than the  limits on the size of extra dimensions obtained using  spectra of hydrogen molecules.
\end{abstract}
	
\maketitle

\section{Introduction}
%It has conventionally understood throughout history that 
Observations indicate that  our universe has three non-compact spatial dimensions, with all forces and particles operating inside these dimensions. However, there are popular theoretical models with extra spatial dimensions such as string theories - see e.g. \cite{Witten1995}. These models have been motivated by search for theory unifying all interactions and producing finite results which do not require hiding infinities using the renormalisation procedures.

%extra spatial dimensions can exist in our universe without disrupting the Standard Model \cite{ADD1998,RS_Model1_1999,RS_Model2_1999}, thus it is important to examine them to determine whether they exist, as this would revolutionise physics. Further motivation lies in the search for a grand unified theory, where popular ideas such as string theory involve extra dimensions - see e.g. \cite{Witten1995}.

In this paper, we examine the Arkani-Hamed - Dimopoulos - Dvali (ADD) model \cite{ADD1998}  which aimed at solving the hierarchy problem  by proposing that gravity can propagate through $n$ extra spatial compact dimensions as well as the regular three non-compact dimensions. Ordinary Newtonian gravitational interaction between elementary particles is many orders of magnitude smaller than other interactions.    In ADD model the observed Newton gravitational  law in our three dimensions will be significantly  strengthened  at distance smaller than the size of extra dimensions $R$.
%(see also Randall-Sundrum models \cite{RS_Model1_1999,RS_Model2_1999}, where the parameter is the distance between the "branes", parallel universes embedded in the multi-dimensional bulk).  
The Newtonian gravitational potential will change from $1/r$ to a more singular $1/r^{n+1}$ dependence at distances smaller than the size of the extra dimensions, $R$. This may be easily explained by the Gauss integrated  flux formula for the gravitational force since size of the "surface" in $(2+n)D$ is proportional to $r^{2+n}$   \cite{ADD1998} (see also the  Randall-Sundrum multi-dimensional models \cite{RS_Model1_1999,RS_Model2_1999} which have similar small-distance gravity). Search for macroscopic effects has not found any deviation from Newton  law. Gravity force has been observed to obey the inverse-square law down to the $\mu$m scale \cite{TanPRL2020}, see also  Refs.~\cite{Kapner2007,Adelberger2007,Chen2016,Vasilakis2009,Terrano2015}. Smaller distances may be probed in atomic phenomena.  Several papers \cite{FengCPL2006,FengIJTP2007,DahiaPRD2016,ZhiGang2008,WangMPLA2013} have attempted to calculate this effect of extra dimensions on spectra of simple atomic systems and placed constraints on $R$ through first order perturbation theory. The problem is that the results  strongly depend on an  unknown (and practically arbitrary) cut-off parameter which has to be introduced to obtain finite energy shift produced by  the gravitational potential $g/r^{n+1}$ for $n>1$. Another problem is that perturabation theory is not applicable if the cut-off parameter is too small and the gravitational potential near the cut-off exceeds the Coulomb potential.  These problems have been solved  in Ref. \cite{Salumbides2015} where the authors  
studied effects of potential $g/r^{n+1}$ between nuclei in H$_2$, D$_2$ and HD$^+$ molecules. Since nuclei in molecules are separated by distance exceeding Bohr radius $a_B$, there is no need in cut-off parameter here.  However, the  effects of the  potential $g/r^{n+1}$  are much bigger for subatomic distances which give the main contribution to the energy shift in atoms. 

In this paper we want to solve the cut-off problem using absence of small size  bound states between two elementary  particles  which could be produced by the singular potential  $g/r^{n+1}$ if the cut-off parameter is too small. Indeed, the highly singular nature of the ADD gravitational potential introduces a fall-to-centre problem, where must exist some new physics mechanism that cuts off the gravitational potential below some cut-off radius $r_c$. To solve the cut-off problem empirically,  we will use simultaneously  two experimental facts: absence of the collapsed gravitational bound states of elementary particles and difference between the measured and calculated (using QED) transition energy. In this way we can get conservative  limits on the size of extra dimensions by excluding the area of unknown physics at distances  smaller than the  electroweak scale.       

\section{Overview of The ADD Model}
The ADD model was first proposed in Ref. \cite{ADD1998} and introduced $n$ extra spatial compactified dimensions of size $R$. There is also a change to the Planck mass, where a higher dimensional Planck mass $M$ is defined with respect to $R$ and replaces the observed three-dimensional Planck mass $M_{Pl}$,
\begin{equation}\label{M}
    \left(\frac{c M_{Pl}}{\hbar}\right)^2 = R^n \left(\frac{c M}{\hbar}\right)^{n+2}.
\end{equation}

This means that Gauss's Law can be applied to the gravitational potential for two masses $m_1$ and $m_2$ separated by distance  $r$:
\begin{eqnarray}\label{e:V_ADD}
\nonumber    
    V(r) = %\begin{dcases}
         -\frac{G m_1 m_2}{r} & \mathrm{for}\,\, r>>R,\\
         -\frac{G m_1 m_2 R^n}{r^{n+1}} = -\frac{\hbar^{n+1} }{c^{n-1} M^{n+2}} \frac{m_1 m_2}{r^{n+1}}& \mathrm{for}\,\, r<<R,
  %  \end{dcases}
\end{eqnarray}
where $G=6.674\time10^{-11}$ m$^3$ kg$^{-1}$ s$^{-2}$ is the regular gravitational constant. The long-range component of the potential  matches the regular Newtonian gravitational potential that is known to be accurate from $\mu$m scales to astronomical scales. However, the short-range component has a more singular form, therefore this is the region where gravitational potential can become significantly stronger with extra dimensions.

For atomic calculations it is convenient to present relation between  Plank mass $M$ for $4+n$ dimensions and radius of extra dimensions $R$ in the following form:  
\begin{equation}\label{e:M}
  Mc^2=(1.22 \cdot 10^{19})^{\frac{2}{n+2}} 10^{-\frac{6n}{n+2}} \left(\frac{3.73a_B}{R}\right)^{\frac{n}{n+2}}\,\,{\rm GeV}.
\end{equation}

We introduce the dimensionless variable $S$ to describe the ratio of the three-dimensional gravitational and Coulomb potential
\begin{equation}
    S \equiv \frac{G m_1 m_2}{e^2}
    %{/4\pi\epsilon_0}
    = \frac{m_1 m_2}{\alpha M_{Pl}^2},
    \label{e:S}
\end{equation}
and therefore the short-range ADD potential can be written as
\begin{equation}\label{VS}
    V(r) = - \frac{\hbar c \alpha S R^n}{r^{n+1}}.
\end{equation}
We can also consider the gravity-Coulomb boundary
\begin{equation}\label{rgc}
    \frac{\hbar c \alpha S R^n}{r_{gc}^{n+1}} = \frac{e^2}{r_{gc}},
 %   4\pi\epsilon_0 r_{gc}},
\end{equation}
which provides the cut-off radius $r_{gc} = S^{1/n} R$ for the applicability of the perturbation  theory for $n>2$. Indeed, for $n>2$  the matrix elements of $V(r)=g/r^{n+1}$ are dominated by the small distance  integral 
 near the cut off radius $r_c$, $\int_b d^3 r/r^{n+1} \sim 1/r_c^{n-2}$. We can present the correction to energy as 
\begin{eqnarray}
\nonumber
 \delta E= <\psi|H|\psi> - <\psi_o|H_0|\psi_0>=\\ 
 \nonumber
 <\psi_0|H-H_0|\psi_0>  + <\psi| H |\psi> - <\psi_0 | H | \psi_0>,
 %=< \psi | H -H_0 |\psi>  +  <\psi | H_0 | \psi> - < \psi_0 | H_0 | \psi_0>,
\end{eqnarray}
where $H-H_0= V(r)$  is the gravitational potential. Since in the area $r<r_{gc}$ the correction  to  the wave function is not small, i.e. condition  $\psi-\psi_0 \ll \psi_0$ is violated for $r_{gc} > r_c$,  the first order perturbation  theory result is incorrect even if $\delta E$ is small.

One may  say that distance $r_{gc}$ may be associated with the scale where all interactions have a comparable strength.

\section{The Perturbation Approach}
As mentioned, we expect  that perturbation theory for the potential $V(r)=g/r^{n+1}$ is applicable for $n<2$ and not applicable for $n>2$. For intermediate case $n=2$ the perturbation theory may provide an estimate with a logarithmic accuracy.  To make a link to ADD theory, we should find estimates for the size of extra dimensions $R_n$, which determines area of highly singular gravitational potential, $V(r)=g/r^{n+1}$ for $r<R_n$. We will do these first estimates using experimental data from muonium, postronium and hydrogen  spectra and perturbation theory.

For $n=1$, we use the known expectation values of $1/r^2$ potentials for $s$-waves \cite{bethe_salpeter_2014} and obtain the energy shift produced by potential Eq. (\ref{VS}): 
\begin{equation}
    \delta E_{1} = \frac{2 \hbar c \alpha S R Z^2}{n_p^3 a^2},
    \label{e:PT_E1}
\end{equation}
where $a$ is the reduced atomic Bohr radius, $Z$ is the nuclear charge (in units of proton electric charge $e$), and $n_p$ is the principal quantum number. 

 For $n=2$ and $n\geq3$, we estimate leading terms enhanced by the small cut off parameter $r_c$ in the first order perturbation energy shift:
\begin{equation}
    \delta E_{2} = \frac{4 \hbar c \alpha S R^2 Z^3}{n_p^3 a^3} \ln\left(\frac{a}{r_c}\right),
    \label{e:PT_E2}
\end{equation}
\begin{equation}
    \delta E_{n\geq3} = \frac{4 \hbar c \alpha S R^n Z^3}{(n-2) n_p^3 a^3 r_c^{n-2}}.
    \label{e:PT_E+}
\end{equation}
We consider two different choices for the cut-off parameter $r_c$.  Firstly, we consider the point $r_{gc}$ where the gravitational and Coulomb potentials are equal.  This choice corresponds to the boundary of applicability of the perturbation theory. 
Defining the variable $\lambda \equiv \frac{1}{n_i^3} - \frac{1}{n_j^3}$ as the inverse cubed principal quantum number difference of the two energy states, we determine the size for any extra dimensions $R_n$ for a given energy shift from gravity for a transition $n_i \rightarrow n_f$, assuming cut-off radius $r_c=r_{gc} = S^{1/n} R$ from equality of the gravitational and Coulomb potentials: 
\begin{equation}
    R_{1} = \frac{a^2 \delta E}{2\hbar c \alpha S Z^2 \lambda},
    \label{e:R1}
\end{equation}
\begin{equation}
    R_{2} = \sqrt{\frac{a^3 \delta E}{4 \hbar c \alpha S Z^3 \lambda \ln(a / S^{1/n} R_2)}},
    \label{e:R2}
\end{equation}
\begin{equation}
    R_{n\geq3} = \sqrt{\frac{a^3 (n-2) \delta E}{4 \hbar c \alpha S^{2/n} Z^3 \lambda}}.
    \label{e:Rlarge}
\end{equation}

Note for $n=2$, $R_2$ must be obtained using iterations; all other $R_n$ can be trivially calculated. To have inputs for $\delta E$, we consider the maximal energy shift between experimental and theoretical energy levels in simple systems. The systems were chosen due to their small deviation between theoretical and experimental results, which is presented in Table \ref{tab:DelE_Inputs}. Spectroscopy results for hydrogen are new, while muonium and positronium measurements are older; see Ref. \cite{MuReview,PsReview} for recent reviews in muonium and positronium spectroscopy. Significant improvements for muonium and positronium $1s-2s$ spectroscopy are expected in the near future \cite{MuMASS,FuturePs1s2s}.

\begin{table}[b]\caption{\label{tab:DelE_Inputs} Input parameters for simple systems. All systems have atomic charge $Z=1$ and reduced Bohr radius $a_B$ (with the exception of positronium with $2a_B$).}
\begin{tabular}{|c|c|c|c|}
    \hline\hline
    System & $\Delta E = E_{exp} - E_{thr}$ & Max. $\Delta E$ (eV) & Ref.\\\hline\hline
    H $1s-2s$ & -0.9(5.4) kHz & $2.2\times10^{-11}$& \cite{Hydrogen1s2s2018}\\
    H $1s-3s$ & 1.2(4.1) kHz & $2.2\times10^{-11}$ & \cite{Hydrogen1s3s2017}\\
    Mu $1s-2s$ & 5.6(9.9) MHz & $6.4\times10^{-8}$ & \cite{Muonium1s2s2000}\\
    Ps $1s-2s$ & -5.78(3.5) MHz & $1.3\times10^{-8}$ & \cite{Ps1s2sThr1,Ps1s2sThr2,Ps1s2sExp}\\\hline\hline
\end{tabular}\end{table}

In our calculations, we use the $S$ values for each system: $S_{ee} = 2.4\times10^{-43}$ for positronium and  $S_{e\mu}=4.9\times10^{-41}$ for muonium. For hydrogen  stable gravitational bound states could be formed between electron and point-like quark. However, when we calculate the gravitational energy shift, the total proton mass density  defines the strength of the gravitational potential. For $n>2$ the quark-electron potential $V(r)=g/r^{n+1}$ with a cut-off $r_c$, which is very small on atomic  scale,  is practically equivalent to the contact potential $V(r)=C \delta(r)$, where $C=\int V(r)d^3 r$. For a finite size proton or nucleus this approximation gives potential  proportional to the mass density $m\rho(r)$, where $\int \rho(r)d^3 r=1$. In light atoms, s-wave electron wave function tends to constant at $r\to 0$ and finite size of the  nucleus has no effect, i.e. we may take $V(r)=C \delta(r)$ for $n>2$. 

For $n=1$ the effect of the potential $V(r)=g/r^2$ is not sensitive to both the cut-off radius and nuclear size. For $n=2$ there is only very weak logarithmic sensitivity to these parameters. Thus we may conclude that formulas presented above may be used to estimate energy shift for hydrogen atom with $S_{ep}=4.4\times10^{-40}$. Use of the proton mass (instead of the quark mass) for the estimate of the cut-off parameter, overestimates  $r_c$ and makes the limits on $R$ weaker but more reliable.  
%\footnote{Instead we could use u-quark charge 2/3  and constituent quark mass $m_q=m_p/3$ to find $r_c$; the current quark mass $m_q \sim 4$ MeV does not include non-mass contributions to gravity.}.

%More rigorous calculations require first to build effective electron-quark and electron-gluon interactions and then  integrate over  the mass density distribution, however it is sufficient to assume a point-like proton in these initial estimates.

A smaller value of the cut-off distance $r_c$ comes from the condition of absence of the gravitational bound states of elementary particles (since they have not been observed experimentally). The estimate follows  from the  equality of the relativistic  kinetic energy $\hbar c/r$ and gravitational energy Eq.~(\ref{e:V_ADD}) for weakly bound states. It leads to
\begin{equation} \label{e:rcR}
 (r_c/R)^n \sim m_1 m_2/M_{Pl}^2, 
 \end{equation}
 which gives  
\begin{equation} \label{e:rc}
r_c \sim \alpha r_{gc}.
\end{equation}
This choice leads to stronger constrains on the size of extra dimensions (see Table  \ref{t:EXD2}). Note that we still use the non-relativistic expression for gravitation interaction (\ref{e:V_ADD}). The estimation of the role of the relativistic effects is presented in the Appendix. Relativistic corrections increase gravitational interaction and make constrain on the radius of extra dimensions $R$ stronger.

\subsection{Deuteron Binding Energy}
Wave function of the deuteron may be found using the short range character of the strong interaction and relatively small binding energy of the deuteron.  Outside the interaction range, we use  solution to the Schrodinger equation for zero potential. Within the interaction range $r_0 = 1.2$ fm,   wave function has a constant value for $s$ orbital.
\begin{equation}\label{e:deut_wavefunc}
    \psi(r) = \left\{\begin{array}{ll}
         \frac{B e^{-\kappa r}}{r} & \text{ for } r>r_0,\\
         \frac{B\, J(0)}{r_0} &  \text{ for } r<r_0,
    \end{array}\right.
\end{equation}
where the normalisation constant $B$ is given by $4\pi B^2 = 2\kappa$ for $\kappa = \sqrt{2m |E|} = 4.56\times10^7$ eV (reduced mass $m = m_p/2$ and binding energy $|E| = 2.22$ MeV). The Jastrow factor, $J(0) = 0.4$ \cite{Jastrow}, is included to account for the nucleon repulsion at short distance.  For $n=1$ we obtain 
\begin{equation}
    \delta E_1 = -\frac{2 \kappa R_1 \alpha S_{pn}}{r_0}, 
\end{equation}
where 
$S_{pn}=0.81 \cdot 10^{-36}$, 
$\alpha S_{pn}=0.59 \cdot 10^{-38}$. Following Ref.\cite{StadnikPRL2018} we take difference between experimental \cite{DeuteronExp1999} and theoretical \cite{DeuteronThr2015} results as $E_{\rm{exp}} - E_{\rm{thr}} = -13.7$ eV. This gives $R_1<3 \cdot 10^{16}$ m.

For $n > 1$ the quark-quark potential $V (r) = -g/r^{n+1}$ with a cut-off $r_c$, which is very small on nuclear  scale, may be approximated by  the contact potential $V (r) = -C\delta(r)$,
where $C = -\int V (r)d^3r$. This contact interaction is similar to the weak interaction mediated by $Z$-boson. From the results of the weak effects calculation we know that finite size of the nucleons produces  suppression of the effects described by the Jastrow factor $J(0)$ \cite{Jastrow}. After accounting for this factor we may assume that nucleons are point-like particles  with interaction $V (r) = -C\delta(r)$. 

Using wave function in Eq. (\ref{e:deut_wavefunc}) we obtain the following estimate for the energy shift in the case of $n>1$: 
\begin{equation}
    \delta E_n = \langle \psi | -C_n \delta(\mathbf{r}) | \psi \rangle = -\frac{ C_n \kappa J(0)^2}{2\pi r_0^2}=-C_n 0.8 \cdot 10^{36} \frac{eV}{m^2},
\end{equation}
where for $n=2$ 
\begin{equation}
C_2= 4 \pi \alpha S_{pn} R_2^2 \ln\frac{r_0}{r_c}, 
\end{equation}
for $n>2$
\begin{equation}
C_n= \frac{4 \pi \alpha S_{pn} R_n^2}{(n-2)} \left(\frac{R_n}{r_c}\right)^{n-2}. 
\end{equation}
We can estimate the cut-off radius $r_c$ from the condition of the absence of the small size gravitational  bound states Eq. (\ref{e:rcR}). Note that gravitational bound state may exist for two quarks while all quarks contribute to the deuteron  binding energy. To have a conservative estimate of the limits on $R$ and $M$,  we use a constituent quark mass  $m_q= 300$ MeV in Eq.~(\ref{e:rcR}) (mass $m_q \sim$ 5 MeV would give smaller $r_c$, bigger energy shifts  and stronger limits). Using $E_{\rm{exp}} - E_{\rm{thr}} = -13.7$ eV we obtain estimates for $R_n$ and $M_n$ presented in Table ~\ref{t:Dt}. 
%{t:R3}.
%We may assume  the cut-off radius for the quark-quark interaction $r_c=r_{gs} = ( S_{qq}/10)^{1/n} R= ( S_{pn}/100)^{1/n} R$ from equality of the gravitational and strong  potential at small distance, with $\alpha_s \approx 10 \alpha$ and the constituent quark mass $m_q\approx m_p/3$.
Formally, these estimates  look like the  strongest constrains among two-body systems. However, deuteron is  a system with the strong interaction and these constrains are probably less reliable than the  constrains from hydrogen, muonium and positronium.

\begin{table}
\caption{\label{t:Dt} Parameters of extra dimensions obtained from the deuteron data.
$n$ is the number of extra dimensions, $R$ is the size of extra dimensions, $r_c$ is the cut-off parameter, $M$ is Planck mass for extra dimensions (\ref{e:M}). 
Numbers in square brackets mean powers of ten.}
\begin{ruledtabular}
\begin{tabular}{cccc}
\multicolumn{1}{c}{$n$}&
\multicolumn{1}{c}{$R$}&
\multicolumn{1}{c}{$r_c$}&
\multicolumn{1}{c}{$M$}\\
&\multicolumn{1}{c}{[m]}&
\multicolumn{1}{c}{[m]}&
\multicolumn{1}{c}{[GeV]}\\
\hline
 2 &  5.1       &  4.5[-19] &  22  \\
 3 &  4.8[-06]  &  9.6[-19] &  25  \\
 4 &  3.8[-09]  &  1.1[-18] &  32 \\
 5 &  5.2[-11]  &  1.2[-18] &  38  \\
 6 &  3.0[-12]  &  1.3[-18] &  43  \\
 7 &  4.0]-13]  &  1.4[-18] &  47  \\
\end{tabular}
\end{ruledtabular}
\end{table}

\section{Numerical calculations}

\begin{table*}
\caption{\label{t:EXD1} Parameters of extra dimensions obtained from comparing experimental and theoretical energy shifts.
Experimental shifts $\Delta E_{\rm expt}$ are taken from Table~\ref{tab:DelE_Inputs}. Parameter $n$ is the number of extra dimensions; 
$\Delta E_a$ is the energy shift given by Eqs.~(\ref{e:PT_E1},\ref{e:PT_E2},\ref{e:PT_E+}); $\Delta E_{\rm PT0}=\langle \psi^{(0)}_{1s}|V|\psi^{(0)}_{ns}\rangle$,
where $\psi^{(0)}$ are unperturbed wave functions; $\Delta E_{\rm PT}=\langle \psi_{1s}|V|\psi_{ns}\rangle$,
where $\psi$ are wave functions obtained by solving Dirac equations including Coulomb and gravitation potentials. Energy shifts $\Delta E_D$, obtained using Dirac equations, 
are fitted to $\Delta E_{\rm expt}$ by varying size of extra dimensions $R$, i.e. $\Delta E_D=\Delta E_{\rm expt}$ . The values of $R_a$ are given by Eqs.~(\ref{e:R1},\ref{e:R2},\ref{e:Rlarge}), $R_D$ 
is chosen to fit experimental energy shift by solving Dirac equations; $r_{gca} = S^{1/n} R_a$, $r_{gc} = S^{1/n} R_D$.
$M$ is Planck mass for extra dimensions (\ref{e:M}). The calculation were done at the cut-off parameter $r_c=r_{gc}$.
Numbers in square brackets mean powers of ten.}
\begin{ruledtabular}
\begin{tabular}{c ccc cccc c}
\multicolumn{1}{c}{$n$}&
\multicolumn{1}{c}{$\Delta E_a$}&
\multicolumn{1}{c}{$\Delta E_{\rm PT0}$}&
\multicolumn{1}{c}{$\Delta E_{\rm PT}$}&
\multicolumn{1}{c}{$R_a$}&
\multicolumn{1}{c}{$r_{gca}$}&
\multicolumn{1}{c}{$R_D$}&
\multicolumn{1}{c}{$r_{gc}$}&
\multicolumn{1}{c}{$M$}\\
&\multicolumn{1}{c}{[eV]}&
\multicolumn{1}{c}{[eV]}&
\multicolumn{1}{c}{[eV]}&
\multicolumn{1}{c}{[m]}&
\multicolumn{1}{c}{[m]}&
\multicolumn{1}{c}{[m]}&
\multicolumn{1}{c}{[m]}&
\multicolumn{1}{c}{[GeV]}\\
\hline
%         write(2,42)n,da*ev,dpt0*ev/nsc,dpt*ev/nsc,re*a0,
%     ,        rc0*a0,re1*a0,rc*a0,pm
\multicolumn{9}{c}{Hydrogen, $1s-2s$, $\Delta E_D=\Delta E_{\rm expt}=2.23\times 10^{-11}$ \ eV}\\
%H 1s-2s  Dexp=  1.861E-11 eV

%  1 &  2.23[-11] &  2.23[-11] &  2.23[-11] &  5.63[+16] &  2.48[-23] &  5.63[+16] &  2.48[-23] &  80.5 \\
%  2 &  2.41[-11] &  2.22[-11] &  2.22[-11] &  3.06[+02] &  6.42[-18] &  3.18[+02] &  6.68[-18] &  2.75 \\
%  3 &  1.69[-11] &  1.69[-11] &  1.69[-11] &  3.37[-04] &  2.56[-17] &  2.93[-04] &  2.23[-17] &  2.15 \\
%  4 &  1.29[-11] &  1.29[-11] &  1.29[-11] &  2.50[-07] &  3.62[-17] &  1.90[-07] &  2.75[-17] &  2.36 \\
%  5 &  1.09[-11] &  1.10[-11] &  1.10[-11] &  3.30[-09] &  4.44[-17] &  2.31[-09] &  3.11[-17] &  2.54 \\
%  6 &  9.44[-12] &  9.45[-12] &  9.45[-12] &  1.86[-10] &  5.13[-17] &  1.21[-10] &  3.33[-17] &  2.70 \\
%  7 &  8.04[-12] &  8.04[-12] &  8.04[-12] &  2.40[-11] &  5.73[-17] &  1.44[-11] &  3.44[-17] &  2.88 \\

  1 &  2.23[-11] &  2.23[-11] &  2.23[-11] &  5.63[+16] &  2.48[-23] &  5.63[+16] &  2.48[-23] &  80.5 \\
  2 &  2.41[-11] &  2.27[-11] &  2.27[-11] &  3.06[+02] &  6.42[-18] &  3.18[+02] &  6.68[-18] &  2.75  \\
  3 &  1.69[-11] &  2.26[-11] &  2.26[-11] &  3.37[-04] &  2.56[-17] &  2.93[-04] &  2.23[-17] &  2.15  \\
  4 &  1.29[-11] &  2.15[-11] &  2.15[-11] &  2.50[-07] &  3.62[-17] &  1.90[-07] &  2.75[-17] &  2.36  \\
  5 &  1.09[-11] &  2.19[-11] &  2.19[-11] &  3.30[-09] &  4.44[-17] &  2.31[-09] &  3.11[-17] &  2.54  \\
  6 &  9.44[-12] &  2.20[-11] &  2.20[-11] &  1.86[-10] &  5.13[-17] &  1.21[-10] &  3.33[-17] &  2.70  \\
  7 &  8.04[-12] &  2.14[-11] &  2.14[-11] &  2.40[-11] &  5.73[-17] &  1.44[-11] &  3.44[-17] &  2.88  \\
\multicolumn{9}{c}{Hydrogen, $1s-3s$, $\Delta E_D=\Delta E_{\rm expt}=2.19\times 10^{-11}$ \ eV}\\
%H 1s-3s  Dexp=  2.192E-11 eV

%  1 &  2.22[-11] &  2.22[-11] &  2.22[-11] &  5.53[+16] &  2.44[-23] &  5.08[+16] &  2.24[-23] &  83.3 \\
%  2 &  2.41[-11] &  2.22[-11] &  2.22[-11] &  3.03[+02] &  6.36[-18] &  3.03[+02] &  6.36[-18] &  2.82 \\
%  3 &  1.66[-11] &  1.66[-11] &  1.66[-11] &  3.34[-04] &  2.54[-17] &  2.77[-04] &  2.11[-17] &  2.22 \\
%  4 &  1.32[-11] &  1.32[-11] &  1.32[-11] &  2.48[-07] &  3.59[-17] &  1.83[-07] &  2.66[-17] &  2.42 \\
%  5 &  1.12[-11] &  1.12[-11] &  1.12[-11] &  3.27[-09] &  4.40[-17] &  2.22[-09] &  2.99[-17] &  2.61 \\
%  6 &  9.58[-12] &  9.58[-12] &  9.58[-12] &  1.84[-10] &  5.08[-17] &  1.16[-10] &  3.20[-17] &  2.79 \\
%  7 &  8.42[-12] &  8.43[-12] &  8.43[-12] &  2.38[-11] &  5.68[-17] &  1.41[-11] &  3.35[-17] &  2.93 \\

  1 &  2.22[-11] &  2.22[-11] &  2.22[-11] &  5.53[+16] &  2.44[-23] &  5.08[+16] &  2.24[-23] &  83.3  \\
  2 &  2.41[-11] &  2.27[-11] &  2.27[-11] &  3.03[+02] &  6.36[-18] &  3.03[+02] &  6.36[-18] &  2.82  \\
  3 &  1.62[-11] &  2.16[-11] &  2.16[-11] &  3.34[-04] &  2.54[-17] &  2.74[-04] &  2.08[-17] &  2.24  \\
  4 &  1.32[-11] &  2.20[-11] &  2.20[-11] &  2.48[-07] &  3.59[-17] &  1.83[-07] &  2.66[-17] &  2.42  \\
  5 &  1.12[-11] &  2.23[-11] &  2.23[-11] &  3.27[-09] &  4.40[-17] &  2.22[-09] &  2.99[-17] &  2.61  \\
  6 &  9.58[-12] &  2.24[-11] &  2.24[-11] &  1.84[-10] &  5.08[-17] &  1.16[-10] &  3.20[-17] &  2.79  \\
  7 &  8.54[-12] &  2.28[-11] &  2.28[-11] &  2.38[-11] &  5.68[-17] &  1.42[-11] &  3.38[-17] &  2.92  \\
\multicolumn{9}{c}{Muonium, $\Delta E_D=\Delta E_{\rm expt}=6.41\times 10^{-8}$ \ eV}\\
%muonium  Dexp=  6.410E-08 eV
%  1 &  6.41[-8] &  6.41[-8] &  6.41[-8] &  1.43[+21] &  7.12[-20] &  1.43[+21] &  7.12[-20] &  2.737 \\
%  2 &  7.04[-8] &  6.28[-8] &  6.28[-8] &  5.67[+04] &  4.00[-16] &  5.95[+04] &  4.20[-16] &  0.201 \\
%  3 &  4.74[-8] &  4.74[-8] &  4.74[-8] &  3.73[-02] &  1.37[-15] &  3.21[-02] &  1.18[-15] &  0.128 \\
%  4 &  3.90[-8] &  3.90[-8] &  3.90[-8] &  2.31[-05] &  1.94[-15] &  1.80[-05] &  1.51[-15] &  0.114 \\
%  5 &  3.32[-8] &  3.32[-8] &  3.32[-8] &  2.73[-07] &  2.38[-15] &  1.97[-07] &  1.71[-15] &  0.106 \\
%  6 &  2.71[-8] &  2.71[-8] &  2.71[-8] &  1.43[-08] &  2.75[-15] &  9.31[-09] &  1.78[-15] &  0.104 \\
%  7 &  2.54[-8] &  2.54[-8] &  2.54[-8] &  1.76[-09] &  3.07[-15] &  1.11[-09] &  1.93[-15] &  0.098 \\

  1 &  6.41[-08] &  6.41[-08] &  6.41[-08] &  1.43[+21] &  7.12[-20] &  1.43[+21] &  7.12[-20] &  2.74  \\
  2 &  7.04[-08] &  6.48[-08] &  6.48[-08] &  5.67[+04] &  4.00[-16] &  5.95[+04] &  4.20[-16] &  0.201  \\
  3 &  4.74[-08] &  6.32[-08] &  6.32[-08] &  3.73[-02] &  1.37[-15] &  3.21[-02] &  1.18[-15] &  0.128  \\
  4 &  3.90[-08] &  6.50[-08] &  6.50[-08] &  2.31[-05] &  1.94[-15] &  1.80[-05] &  1.51[-15] &  0.114  \\
  5 &  3.32[-08] &  6.65[-08] &  6.65[-08] &  2.73[-07] &  2.38[-15] &  1.97[-07] &  1.71[-15] &  0.106  \\
  6 &  2.71[-08] &  6.32[-08] &  6.32[-08] &  1.43[-08] &  2.75[-15] &  9.31[-09] &  1.78[-15] &  0.104  \\
  7 &  2.54[-08] &  6.78[-08] &  6.79[-08] &  1.76[-09] &  3.07[-15] &  1.11[-09] &  1.93[-15] &  0.098 \\
\multicolumn{9}{c}{Ps, $1s-2s$, $\Delta E_D=\Delta E_{\rm expt}=1.34\times 10^{-8}$ \ eV}\\
%Ps       Dexp=  2.978E-09 eV

  1 &  1.34[-8] &  1.34[-8] &  1.34[-8] &  2.49[+23] &  5.97[-20] &  2.49[+23] &  5.97[-20] &  4.9[-1] \\
  2 &  1.50[-8] &  1.26[-8] &  1.26[-8] &  1.04[+06] &  5.08[-16] &  1.10[+06] &  5.39[-16] &  4.7[-2] \\
  3 &  9.71[-9] &  9.71[-9] &  9.71[-9] &  2.86[-01] &  1.78[-15] &  2.43[-01] &  1.51[-15] &  3.8-2] \\
  4 &  7.97[-9] &  7.97[-9] &  7.97[-9] &  1.14[-04] &  2.51[-15] &  8.75[-05] &  1.94[-15] &  4.0[-2] \\
  5 &  6.58[-9] &  6.59[-9] &  6.59[-9] &  1.03[-06] &  3.08[-15] &  7.20[-07] &  2.16[-15] &  4.2[-2] \\
  6 &  5.68[-9] &  5.68[-9] &  5.68[-9] &  4.51[-08] &  3.56[-15] &  2.93[-08] &  2.31[-15] &  4.4[-2] \\
  7 &  5.16[-9] &  5.17[-9] &  5.17[-9] &  4.87[-09] &  3.97[-15] &  3.02[-09] &  2.46[-15] &  4.5[-2] \\
\end{tabular}
%\footnotetext[1]{Electron loop contribution only.}
\end{ruledtabular}
\end{table*}

\begin{table*}
\caption{\label{t:EXD2} The same as Table~\ref{t:EXD1} but
the calculations were done at the smaller values of the cut-off parameter $r_c=r_{gca}/100$, chosen from the condition of the absence of the gravitational  bound states of elementary particles.
%Numbers in square brackets mean powers of ten.
}
\begin{ruledtabular}
\begin{tabular}{c ccc ccc c}
\multicolumn{1}{c}{$n$}&
\multicolumn{1}{c}{$\Delta E_a$}&
\multicolumn{1}{c}{$\Delta E_{\rm PT0}$}&
\multicolumn{1}{c}{$\Delta E_{\rm PT}$}&
\multicolumn{1}{c}{$R_a$}&
\multicolumn{1}{c}{$r_{c}$}&
\multicolumn{1}{c}{$R_D$}&
%\multicolumn{1}{c}{$r_{gc}$}&
\multicolumn{1}{c}{$M$}\\
&\multicolumn{1}{c}{[eV]}&
\multicolumn{1}{c}{[eV]}&
\multicolumn{1}{c}{[eV]}&
\multicolumn{1}{c}{[m]}&
\multicolumn{1}{c}{[m]}&
\multicolumn{1}{c}{[m]}&
%\multicolumn{1}{c}{[m]}&
\multicolumn{1}{c}{[GeV]}\\
\hline
%         write(2,42)n,da*ev,dpt0*ev/nsc,dpt*ev/nsc,re*a0,
%     ,        rc0*a0,re1*a0,rc*a0,pm
\multicolumn{8}{c}{Hydrogen, $1s-2s$, $\Delta E_D=\Delta E_{\rm expt}=2.23\times 10^{-11}$ \ eV}\\
  2 &  1.01[-12] &  9.49[-13] &  1.26[-08] &  3.06[+02] &  6.42[-20] &  5.73[+01] &  6.5 \\
  3 &  7.58[-13] &  7.59[-13] &  8.41[-10] &  3.37[-04] &  2.56[-19] &  2.35[-05] &  9.8 \\
  4 &  7.85[-13] &  7.86[-13] &  3.23[-10] &  2.50[-07] &  3.62[-19] &  1.08[-08] &  16 \\
  5 &  8.04[-13] &  8.05[-13] &  1.96[-10] &  3.30[-09] &  4.44[-19] &  1.07[-10] &  23 \\
  6 &  8.17[-13] &  8.17[-13] &  1.36[-10] &  1.86[-10] &  5.13[-19] &  4.97[-12] &  30 \\
  7 &  8.25[-13] &  8.26[-13] &  9.73[-11] &  2.40[-11] &  5.73[-19] &  5.59[-13] &  36 \\
\multicolumn{8}{c}{Hydrogen, $1s-3s$, $\Delta E_D=\Delta E_{\rm expt}=2.19\times 10^{-11}$ \ eV}\\
  2 &  1.10[-12] &  1.03[-12] &  1.33[-08] &  3.03[+02] &  6.36[-20] &  5.69[+01] &  6.5 \\
  3 &  8.18[-13] &  8.19[-13] &  7.21[-10] &  3.34[-04] &  2.54[-19] &  2.33[-05] &  9.8 \\
  4 &  8.48[-13] &  8.49[-13] &  3.49[-10] &  2.48[-07] &  3.59[-19] &  1.07[-08] &  16 \\
  5 &  8.61[-13] &  8.62[-13] &  1.09[-10] &  3.27[-09] &  4.40[-19] &  1.06[-10] &  23 \\
  6 &  8.74[-13] &  8.75[-13] &  7.92[-11] &  1.84[-10] &  5.08[-19] &  4.91[-12] &  30 \\
  7 &  8.85[-13] &  8.86[-13] &  7.33[-11] &  2.38[-11] &  5.68[-19] &  5.53[-13] &  36 \\
  \multicolumn{8}{c}{Muonium,  $\Delta E_D=\Delta E_{\rm expt}=6.41\times 10^{-8}$ \ eV}\\
  2 &  2.84[-09] &  2.62[-09] &  2.24[-05] &  5.67[+04] &  4.00[-18] &  1.01[+04] &  0.49 \\
  3 &  2.17[-09] &  2.17[-09] &  1.55[-06] &  3.73[-02] &  1.37[-17] &  2.60[-03] &  0.58 \\
  4 &  2.25[-09] &  2.25[-09] &  9.50[-07] &  2.31[-05] &  1.94[-17] &  1.00[-06] &  0.78 \\
  5 &  2.32[-09] &  2.33[-09] &  1.60[-06] &  2.73[-07] &  2.38[-17] &  8.89[-09] &  0.97 \\
  6 &  2.35[-09] &  2.35[-09] &  4.77[-07] &  1.43[-08] &  2.75[-17] &  3.83[-10] &  1.14 \\
  7 &  2.37[-09] &  2.37[-09] &  2.82[-07] &  1.76[-09] &  3.07[-17] &  4.09[-11] &  1.28 \\
\multicolumn{8}{c}{Ps, $1s-2s$,$\Delta E_D=\Delta E_{\rm expt}=1.34\times 10^{-8}$ \ eV}\\
  2 &  5.89[-10] &  5.21[-10] &  6.02[-06] &  1.04[+06] &  5.08[-18] &  1.85[+05] &  0.11 \\
  3 &  4.55[-10] &  4.55[-10] &  3.27[-07] &  2.86[-01] &  1.78[-17] &  1.99[-02] &  0.17 \\
  4 &  4.72[-10] &  4.73[-10] &  2.00[-07] &  1.14[-04] &  2.51[-17] &  4.92[-06] &  0.27 \\
  5 &  4.81[-10] &  4.82[-10] &  7.83[-08] &  1.03[-06] &  3.08[-17] &  3.34[-08] &  0.38 \\
  6 &  4.91[-10] &  4.92[-10] &  8.26[-08] &  4.51[-08] &  3.56[-17] &  1.21[-09] &  0.48 \\
  7 &  4.93[-10] &  4.94[-10] &  4.12[-08] &  4.87[-09] &  3.97[-17] &  1.13[-10] &  0.58 \\

\end{tabular}
%\footnotetext[1]{Electron loop contribution only.}
\end{ruledtabular}
\end{table*}

We performed the numerical calculations of the energy shifts caused by gravitation potential in order to find limits on the size of extra dimensions $R$.
We do this in two ways. First, we use perturbation theory (PT) approach described in previous section. Second, we solve Dirac equations for the $1s$, $2s$, $3s$ states of hydrogen, muonium and positronium. The energy shift is found as a difference of  energies calculated in the Coulomb potential and in the potential with added gravitational contribution. The size of  extra dimensions $R$ is used in this approach as a fitting parameter to fit input energy shifts from Table~\ref{tab:DelE_Inputs} (we call these shifts "experimental" shifts and use notation $\Delta E_{\rm expt}$ for them). The calculations with Dirac equations are done to check whether PT really works.
Note that small value of the correction to the energy does not necessary mean that PT is applicable. The correction is small because it comes from a very small region in the vicinity of the cut-off radius. However, in this region the change of the potential is not small leading to large change in the wave function and breaking of PT.

We perform the calculations using two ways of defining the cut-off parameter $r_c$ ($V_g(r)=V_g(r_c)$ for $r <r_c$).
Using $r_c = r_{gc}$ (see Eq. (\ref{rgc})) leads to conservative estimates of the size of extra dimensions which are in agreement with the perturbation theory calculations. Corresponding results are presented in Table~\ref{t:EXD1}.
In another approach we use much smaller value of $r_c$ which we find from the condition of the absence of the  small size gravitational bound states of elementary particles. 
% (since they have not been  observed experimentally).
 In practice this means that the wave function, found from solving the Dirac equation, does not oscillate at small distances.
We found that in all cases $r_c = r_{gca}/100$, where $r_{gca}=S^{1/n}R_a$,  in excellent agreement with an estimate $r_c \sim \alpha r_{gc}$ from Eqs. (\ref{e:rcR},\ref{e:rc}). 
The results are presented in Table~\ref{t:EXD2}.

In Tables \ref{t:EXD1} and \ref{t:EXD2} we present energy shifts and corresponding values of $r_c$, $R$ and $M$ obtained in several different ways.
$\Delta E_a$ is calculated using formulae (\ref{e:PT_E1},\ref{e:PT_E2},\ref{e:PT_E+}) but with the radius of extra dimensions found from the fitting of the energy shift $\Delta E_D$ obtained by  solving the Dirac equations. 
$\Delta E_{\rm PT0}=\langle \psi^{(0)}_{1s}|V|\psi^{(0)}_{ns}\rangle$, where $\psi^{(0)}$ are unperturbed wave functions. 

To avoid a misunderstanding, we should note that we used  slightly different method of cut-off in the analytical approach and in the numerical solution of the Dirac equation. 
$\Delta E_{\rm PT0} = \Delta E_a$ if integration for $\Delta E_{\rm PT0}$ is done from $r_c$ to infinity, as in the analytical approach. We did such calculations as a test and found good agreement between the two. However, in the tables we present values of $\Delta E_{\rm PT0}$ and $\Delta E_{\rm PT}$ found by integrating from zero to infinity.
This is done to make meaningful comparison with the energy shift obtained from solving the Dirac equations. The same gravitation potential is used to calculate $\Delta E_{\rm PT0}$, $\Delta E_{\rm PT}$ and $\Delta E_D$ in the Dirac equations.
If $\Delta E_{PT0} = \Delta E_{PT}=\Delta E_{\rm expt}$ ( we have  $\Delta E_D=\Delta E_{\rm expt}$)
then one could say that PT works. We see that this is the case only for $r_c=r_{gca}$ (Table~\ref{t:EXD1}).
%The general trend suggests, that the higher the number of extra dimensions $n$, the larger the difference between PT and solving the Dirac equations.  

The values of $R_a$ and $r_{gca}$ are given by Eqs.~(\ref{e:R1},\ref{e:R2},\ref{e:Rlarge}) and (\ref{rgc}). The values of $R_D$ and $r_{gc}$ are found from solving the Dirac equations. 
%Here again the difference between values of $R_a$ and $R_D$  illustrates poor accuracy  of  PT.
 We believe that the results obtained with the use of Dirac equations are more reliable.
The Planck mass for extra dimensions $M$ was found using the Dirac equations value of $R$ ($R_D$).

%Table~\ref{t:R3} summarise the strongest  limits on the parameters of extra dimensions which come from solving the Dirac equations for the $1s - 2s$ transition in hydrogen for $r_c=r_{gc}$ and $r_c=r_{gca}/100$. The table also includes the results for deuterium obtained with the equations of previous section. 

\section{Conclusion} 
In the present paper we investigated possibilities  to probe ADD gravitational potential Eq. (\ref{e:V_ADD})  at sub-atomic distances. This potential does not include relativistic corrections. However, as it is demonstrated in Appendix, relativistic corrections increase gravitational interaction and  lead to  much stronger  constrains. Therefore, our estimates of the limits on the size of extra dimensions, based on  the potential Eq. (\ref{e:V_ADD}),  are conservative. By imposing the condition that there are no gravitational bound states between  stable elementary particles (which have not been observed), we cut-off  area of of unknown physics at distances  smaller than the  electroweak scale.   Constrains on the size of extra dimensions $R$  and Plank mass $M$, obtained using the difference between the experimental and calculated value of the  deuteron binding energy, have been presented in the Table \ref{t:Dt}. 
However, deuteron is  a system with the strong interaction and these constrains are probably less reliable than the  constrains from hydrogen, muonium and positronium presented in the Table ~\ref{t:EXD2}. The perturbation theory is not applicable if the main contribution to the energy shift comes from the area where the gravitational potential exceeds the Coulomb potential. Therefore, we obtain the energy shift using solution of the Dirac equation including both Coulomb and gravitational potential. Corresponding limit on the size of extra dimensions is denoted by $R_D$.

\section*{Acknowledgements}
 This work was supported by the Australian Research Council Grants No. DP190100974 and DP200100150.
 
 \section{Appendix: relativistic estimates}
 
 As known, all types of energy contribute to the gravitational potential. For example, kinetic energy of nearly massless quarks ($m_u \approx $ 3 MeV, $m_d \approx$ 5 MeV) give significant contribution to the nucleon mass. Kinetic energy of ultra-relativistic particle may be estimated as $E_k=pc \sim \hbar c/r$, so mass $m$ in the gravitational  potential should be replaced by $\hbar /cr$, and corresponding gravitational interaction energy of two particles is   
\begin{equation}\label{Urel}
U \sim - \frac{G}{r}\left(\frac{\hbar}{cr}\right)^2 \left(\frac{R_n}{r}\right)^n
\end{equation}
For weakly bound states kinetic energy and potential energy compensate each  other, $E_k=|U|$, and this equality defines the  cut-off radius excluding formation of the bound states, $r_c=\hbar/M_n c$, where $M_n$ is the Plank mass for extra $n$ dimensions  -see Eq. (\ref{M}). Note that one can find similar estimates in Ref. \cite{DeepBoundStatesPaper}.

Perturbation theory gives the following estimate for the energy shift produced by the potential Eq. (\ref{Urel}) with the cut-off $r_c=\hbar/M_n c$ (in natural units $\hbar=c=1$):
\begin{equation}\label{UrelDeltaE}
\delta E \sim \frac{ 4 \pi \psi(0)^2}{n M_n^2} 
\end{equation}
For hydrogen $\delta E < 2.2 \cdot 10^{-11} eV$ and we obtain the  limit $M_n > $ 100 GeV $/n^{1/2}$, which is much stronger than the limit obtain using the non-relativistic gravitational potential. For deuterium the limit is even stronger, $M_n > $ 170 GeV $/n^{1/2}$. 

Thus, we conclude that use of the non-relativistic gravitational potential underestimates the energy shift  $\delta E$ and  gives conservative limits on the Plank mass $M_n$ and radius of extra dimensions $R_n$.  
   
%\bibliographystyle{apsrev}
%\bibliography{biblio}

%\end{document}

\end{document}